# Programmable Turbine Failsafe System for Pico-Hydroelectric Power in the Nepal Himalayas

James Hsi-Jen Yeh, Rick Sturdivant, Member, IEEE,
Dept. of Eng. and Computer Science
Azusa Pacific University
901 E Alosta Ave, Azusa, CA 91702

Mark Stambaugh, Alex Zahnd
RIDS-USA/RIDS-Switzerland

*Abstract*—In this paper, we describe a novel turbine failsafe system designed for a pico-hydroelectric power generation. We designed and built a pico-hydroelectric system in a remote rural Himalayan village. We developed a Prioritized Load Control System (PLCS) that diverts excess electrical power to useful purposes such as heating up bath water and the air in the community center when the electrical demand from the village is low; instead of dumping the excess electricity back into the river in the form of heat. A critical part of the PLCS system is the failsafe mechanism where the turbines and electrical components on the grid are protected in case any part of the PLCS system stops functioning, possibly intermittently. Our novel failsafe system contains the following enhancements compared to previous systems: (1) adjustable threshold voltage, (2) controllable fractional power diversion with adjustable parameters, (3) automatic reset with adjustable parameters. In addition, the failsafe subsystem uses the widely available Arduino platform and programming environment, and JSON for human readable and writable communications, demonstrating their suitability for critical applications. (*Abstract*)

I. INTRODUCTION

As a central part of the pilot pico-hydroelectric project in the remote Himalayan village of Moharigaun in the Jumla district of Nepal [1-3], we developed a Prioritized Dump Load Control (PLCS) system, to fully utilize the excess electrical power when the electrical demand from the village is low. When the electrical usage falls, for example during the night, the speed and voltage of the hydroelectric turbine generator will rise, causing damage to the turbines and connected electrical devices [4]. A dump load circuit is typically used to heat up resistors and dump the excess electricity back to the river. The purpose of the PLCS is to create value for the villagers with the excess power, such as heating up water tanks for shower water and biogas digesters. The users of the system can set their own priority, based on their needs, regarding the distribution of the excess power, in such a way that low priority tanks are given power only when high priority tanks have reached their target temperature. The AC Load Controller subsystem of the PLCS, which provides computer-controlled power to multiple useful loads dispersed throughout the village as excess power becomes available, is described in a separate article [5].

As a part of the hydroelectric generation system, a failsafe is installed close to the turbine to protect the turbines and the electric components connected to the grid. Previously, a crowbar circuit is often installed to protect the turbine from damage if the turbine speed and voltages rise uncontrollably [6] due to a reduction in electrical load. However, the crowbar circuit requires manual reset. Also, when it is activated, power generation is interrupted.

Previous efforts have been developed to provide electronic control of the dump load using programmable logic controllers and microcomputers [7, 8]. We designed and created a new failsafe subsystem that aims to achieve the following:

1) Threshold voltages that can be adjusted locally or remotely.
2) Power dumping that is a fraction of the generated power, which is a (linear, proportional-integral-derivative, or non-linear) function of the voltage above the threshold, with parameters and functions adjustable and selectable locally or remotely. This has the potential of smoothing out the voltage fluctuation in the turbine and in the grid, while continuing supplying power to the grid.
3) Automatic reset via fractional dump power back-off to mitigate the need for manual reset, with parameters adjustable locally or remotely. This additionally smooth out potential voltage fluctuations.

As a final safeguard, a crowbar circuit is additionally deployed in case the failsafe subsystem fails.

This paper describes the novel failsafe subsystem, which is based on the Arduino platform and programmed using the Arduino Integrated Development Environment, and uses JSON (JavaScript Object Notation) for communications.

## II. FAILSAFE HARDWARE

### A. Platform

For this project, we chose the Arduino Mega platform. The Arduino Mega is a widely available board based on the Atmel ATMega2560 microcontroller [9]. It is low cost, and includes 54 digital I/O pins, out of which 15 can simulate analog output via PWM (pulse-width modulation), and 16 analog input pins, which are required for this project. It contains 256kB of flash memory, 8kB of SRAM and runs at 16MHz. The Arduino Integrated Development Environment (IDE) is one of the most popular open-source development systems. It is extremely stable and is very well documented and supported, with open-source user-contributed libraries for many common task and specialized components.

TABLE I.   FAILSAFE CARRIER BOARD SUBSYSTEMS

| Item | Qty | Main Component | Primary Function |
|---|---|---|---|
| 1 | 1 | ADM2481 | RS485 Communications |
| 2 | 1 | PQMC3 7805-1000 | On-Board Power |
| 3 | 3 | passive | Voltage Sensor |
| 4 | 3 | STGWT40 | Dump Load Drivers |
| 5 | 3 | LED | LED Indicator |
| 6 | 1 | passive | 1-Wire Temperature Sensor |
| 7 | 3 | passive | Current Sensor |
| 8 | 1 | passive | 4-20mA Sensor Interface |

The Arduino-compatible systems have very fast boot time, on the order of < 5 seconds, as compared to other systems such as the Raspberry Pi, where the boot time can be 30 seconds or more. In critical power control applications, slow boot times after a power failure or system crash can be disastrous. In addition, the Arduino system is essentially a single-threaded execution environment. Therefore, all program control flow and critical timing are in control of the programmer. This is as opposed to platforms running an operating system such as Linux, where the operating system and services can introduce complexity and timing uncertainty and instability to critical operations.

A carrier board was designed and fabricated with all of the external components, including communications chips, control chips, and sensing chips, on which the Arduino Mega board mates directly via its header pins. This created a mechanically and electrically stable interface. The subsystems on the carrier board are listed in Table 1.

### B. Dump Load Drivers

The dump load uses high-voltage STGWT40 IGBT's (Integrated Gate Bipolar Transistor) to divert excess power to each of the 3 banks of dump resistors. The IGBT's can handle voltages up to 650V and can dump up to 40A of current. It has a very fast turn-on time of 40ns and a turn-off time of 142ns.

The IGBT's require a turn-on voltage of 15V, which is supported by the IX4427N driver chips that boost the 0-5V Arduino Mega output to the 0-15V gate drive voltages.

The dump load for each pair of turbines consists of two 230VAC 2000W water heater elements in series to handle the 300VDC output of the turbines. They are immersed in the turbine exhaust water, and the excess heat is dissipated back into the river. No effort was made to utilize this energy because the PLCS' normal mechanisms used to utilize excess energy are expected to consistently function correctly.

### C. Communications

For communications, we chose the RS485 interface, which is capable of half-duplex communications over twisted pair for up to 1.2km, and can connect up to 32 devices. RF communications were not used due to licensing issues in Nepal, and the lack of line-of-sight path between the turbine house and the control system in the village. The Arduino Mega supports RS485 with a few hardware components, and within its core libraries.

### D. Sensors

The carrier board contains various passive components and connectors for voltage and current measurements of the three turbine input channels. These inputs, scaled and limited to 0V to 5V, are fed into the analog input channels of the Arduino Mega.

The carrier board also provides connectors for the 1-Wire interface, used for the DS18B20 temperature sensors, which can be connected in parallel. In addition, it provides circuitry for industry standard 4-20mA pressure sensors.

## III. FAILSAFE SOFTWARE

### A. Fractional Power Diversion

When an over-voltage situation occurs, we designed the failsafe subsystem to dump a fraction of the generated power. Our implementation outputs a PWM (pulse-width modulation) signal to the IGBT to dump a percentage of the generated power. The PWM output

duty cycle is currently linear with respect to the amount of over-voltage

$$PWM_i = \frac{V_i - V_{i,th}}{V_{i,range}}$$

if $V_i > V_{i,th}$
subject to the constraint $0\% \leq PWM_i \leq 100\%$

where
$PWM_i$ is the PWM output for channel $i$
$V_i$ is the most recent voltage sample for channel $i$
$V_{i,th}$ is the threshold voltage for channel $i$
$V_{i,range}$ is the voltage range over which the PWM goes to its max duty cycle for channel $i$

Both $V_{i,th}$ and $V_{i,range}$ can be set by the user as parameters through the RS485 interface via JSON commands. Proportional-Integral-Derivative (PID) control or non-linear fraction power diversion utilizing the same PWM mechanism can be employed. Multiple algorithms can be placed into the code, and selected via user command or automatically by the system as conditions change.

### B. Software Reset and Back-Off

The current implementation utilizes the same linear PWM algorithm to back off the fraction of dumped power as the turbine voltage decrease. This mitigates the need for manual reset. To avoid rapid changes in the PWM cycle due to voltage noise and fluctuations, hysteresis was implemented in software for each of the PWM channels. Specifically, a hold time $T_{i,hold}$ is set for each voltage sensing channel.

If the most recent over-voltage sample is greater than the previous over-voltage sample for the same channel, then the PWM duty cycle for the channel is calculated and applied immediately. This is designed to provide immediate protection for over-voltage conditions.

However, if the most recent over-voltage sample is less than the previous over-voltage sample for the same channel, the PWM duty cycle for the channel will be calculated and applied only after the expiry of the hold time (for that channel).

The PWM frequency is set to 1kHz, so there is a 1ms period over which for part of the cycle the IGBT is on, and the other part when the IGBT is off. At the on-part of the cycle, the voltage may drop dramatically. Since the voltage sampling period (200μs) is much shorter than the PWM period, changing the PWM duty cycle before the completion of a PWM period will lead to wild fluctuation in voltage and/or cause the Arduino Mega PWM logic to malfunction. In addition, even upon the completion of a PWM period, we may want to hold the PWM duty cycle to stabilize the turbine voltage.

The default hold times are set to 100ms; they can be adjusted independently for each channel via the JSON interface.

Similar to the Power Diversion algorithm described above, PID and/or non-linear back-algorithms can be coded and deployed.

### C. Critical Timing and Optimization

The smart priority dump load controller (PLCS) required the simultaneous monitoring of three (3) voltages form the charge controllers. Initially, the voltage sampling period was set at 100ms for each of the 3 channels. The three voltages are sampled in a round robin manner - each one in turn. So, the ADC samples at a period of 100ms / 3 = 33ms.

In the absence of electrical load, the turbine voltage can increase very quickly. During testing, this was found to be inadequate, and the physical crowbar circuit. Refined calculation suggests the voltage can increase at the rate of 5833V/s or ~6V/ms. This means that the voltage can go 200V above the operating point before the first dump load turns on. Therefore, a much lower sampling period (higher sampling rate) of per channel was required. For example, using a 2ms sampling period, the voltage can rise 12V between samples.

Aggressive optimization of the code was performed to dramatically decrease the sampling period to 200μs for each of the 3 channels, or 67μs ADC sample period. This is achieved by (1) programming the on-chip ADC directly to reduce the conversion time, (2) using hardware interrupt when conversion is complete, and (3) converting divisions into multiplications followed by bit shifting.

The Atmel Mega 2560 chip's ADC contains a pre-scaler which determines the sample conversion period. The default pre-scaler of 128 chosen the in the Arduino core library resulted in an acquisition time of 100μs. However, the pre-scaler can be changed to decrease the sample acquisition time at the expense of reduced accuracy [10]. For the current project, it was found that the accuracy will not be significantly impacted by choosing a pre-scaler of 16, which resulted in an ADC sample acquisition time of 13μs. This allowed for the attainment of 200μs for each of the 3 channels, or 67μs ADC sample period.

The default Arduino core library provides the `analogRead()` function to access the ADC. This is a blocking operation meaning that all other execution in the single-threaded environment is suspended pending the completion of the sample acquisition. The use of interrupts upon the completion of the sample acquisition allows the processor to complete other tasks such as process incoming serial commands.

Because the failsafe algorithm is executed whenever the voltage sample exceeds the threshold, the amount of time for its execution is critical. The Atmel Mega 2640 has native support for integer multiplication, but not division. We found that the arithmetic division by $V_{i,range}$ executes in excess of 100us, which is greater than the ADC sampling period. To optimize the process, we note that the $V_{i,range}$ only change when they are set via RS485 or local console. This does not occur frequently compared to ADC sampling. Therefore, scale values $S_i = 256 / V_{i,range}$ is computed when the user (infrequently) sets a new $V_{i,range}$. The (frequent) computation of the PWM duty cycle then becomes

$$PWM_i = \frac{S_i(V_i - V_{i,th})}{256}$$

where the division by 256 is replaced by a very fast right shift of 8 bits.

The execution speed constraint and optimization must also be observed for any additional (more complicated) PID or non-linear algorithm that is programmed and deployed into the failsafe subsystem.

### D. Command Protocol

The command protocol was chosen to use the JSON (JavaScript Object Notation) format, which is widely used in Internet communications applications [11]. It is beneficial because it is human readable and writable, and stable exiting library exists for the Arduino to parse JSON. While it uses more bytes than competing binary formats such as MODBUS, we decided that the ability for human operators to easily issue commands and understand responses is critical for this application, which is located in a very remote area.

The code is written in a way so the list of commands can be easily extended by simply adding entries into a command table, which is composed of a pair of data: a JSON command string, and its associated function to call. The associated function will be given as parameters (1) the input (command) JSON object that it acts on, and (2) the output (response) JSON object that it modifies. The output JSON object will be sent as response to the requester.

Commands can be entered via the RS485 interface or a serial console attached the USB interface of the Arduino Mega. Both use the same JSON command format listed below. For commands over the RS485 interface, to detect data corruption due to noise, 16-bit CRC (cyclic-redundancy-code) is calculated and transmitted after the command message; it is checked by the receiver for error detection. This is implemented in software via an FSM (finite-state machine). The CRC is not used for local serial console. A serial timeout of 5 seconds is implemented, in case the transmitter or the communications link fails mid-message.

### E. Command and Response JSON Format

Because a linearly chained RS485 interface is used, which is in effect a shared media among all of the connected devices, each JSON command sent from the controller needs to identify its target device. Therefore, each device is assigned an identifier number. In the case of the failsafe board, it is assigned an identifier of 1. The JSON command addressing the dump controller starts with the following JSON fields, which can be shortened if required:
```
"requestId":1
"manufacturerName":"RDIS"
"modelName":"Failsafe"
```

Similarly, the JSON response from the dump controller requires self-identification, and start with the following JSON fields:
```
"responseId":1
```
Current supported commands and sample responses are listed in Table 2.

TABLE II. FAILSAFE JSON COMMANDS AND SAMPLE RESPONSES

| |
|---|
| Command (retrieves the PLCS status): <br> `{"requestId":1,"manufacturerName":"RDIS","modelName":"Failsafe","getStatus":null}` |
| Response (reports time in milliseconds since restart, current sensor, pressure sensor, temperature sensor(s), voltage sensors, status of the dump controller): <br> `{"responseId":1,"upTime":55247,"current":509,"pressure":0,"temp":[23],"voltage":[0,0,0],"dump":[0,0,0]}` |
| Command (retrieves the internal parameters): <br> `{"requestId":1,"manufacturerName":"RDIS","modelName":"Failsafe","getParameter":null}` |
| Response (reports $V_{i,th}$, $V_{i,range}$, $T_{i,hold}$ described below): <br> `{"responseId":1,"V_th":[128,128,128],"V_range":[118,118,118],"T_hold":[60,60,60]}` |
| Command (retrieves the 1-wire address of the attached DS18b20 temperature probes): <br> `{"requestId":1,"manufacturerName":"RDIS","modelName":"Failsafe","getTempAddress":null}` |
| Response: <br> `{"responseId":1,"tempAddr":["28:98:2C:77:91:06:02:6C"]}` |

| | |
|---|---|
| Command (set $V_{i,range}$ - the voltage range which the PWM varies for the $i$th channel): | |
| `{"requestId":1,"manufacturerName":"RDIS","modelName":"Failsafe","setRange":[1,90]}` | |
| Response: | |
| `{"responseId":1,"V_range":[1,90]}` | |
| Command (set $V_{i,th}$ - the threshold voltage where the PWM starts for the $i$th channel): | |
| `{"requestId":1,"manufacturerName":"RDIS","modelName":"Failsafe","setThreshold":[0,120]}` | |
| Response: | |
| `{"responseId":1,"V_th":[0,120]}` | |
| Command (set the hysteresis hold time $T_{i,hold}$ for the $i$th channel): | |
| `{"requestId":1,"manufacturerName":"RDIS","modelName":"Failsafe","setHoldTime":[2,50]}` | |
| Response: | |
| `{"responseId":1,"T_hold":[2,50]}` | |
| Command (reset the system): | |
| `{"requestId":1,"manufacturerName":"RDIS","modelName":"Failsafe","RESET":null}` | |
| Response: | |
| No response - system reset | |

*F. Diagnostics and Statistics Reporting*

The three-color LED's on the carrier board are used to indicate the status of the PLCS to the operator in an intuitive manner. The green LED flashes with any activity on the RS485 or local serial console.

The blue and red LED work in conjunction to indicate the status of each of the dump channels. The blue LED blinks, once, twice, or thrice to identify which of the 3 channels are being reported (by the red LED). The intensity of the red LED indicates the duty cycle of the PWM for the channel identified by the blue LED.

Extensive diagnostic and statistics reporting can be enabled by compiler directives (`#define`'s). If enabled, the status of the PLCS system will be collected and printed on the local serial console every 2 seconds. Also, statistics on how many times each of the voltage channels have been sampled and statistics on the voltages themselves can be collected and printed.

*G. Temperature Sensing*

Multiple DS18B20 temperature sensors can be connected to the 1-wire interface. At boot time the connected temperature sensors are queried and their 1-wire addresses stored. Each of the temperature sensors is read with a period of 30 seconds. They are also read in a round robin manner. The sample acquisition time for each sensor is 750ms, which is very long. So non-blocking sample acquisition is implemented.

IV. RESULTS

The PLCS has been installed in the village of Moharigaun in the Jumla district of Nepal, located in the remote Himalayas. In November 2018, a trip was made to install the PLCS (Figure 1). It was during this installation that it was discovered that the 100ms voltage sample time was insufficient to control the voltage of the turbine. The crowbar circuit would trigger before the PLCS triggers, causing disruption in power and requiring manual reset.

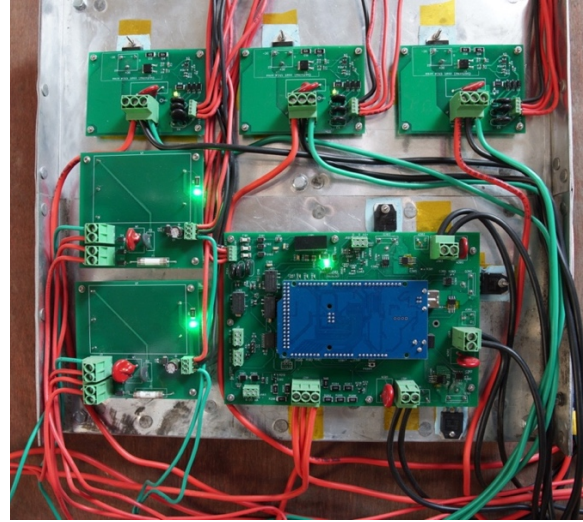

Fig. 1. Image of the Arduino Mega board (blue) mounted on the Failsafe carrier board (green, lower right)

The critical timing and optimization were performed in the control software, and they significantly reduce the sample time. In April 2019, another trip to Moharigaun was made to install the new software. In addition, optimization was made to the charge controller to hold the voltage at 300VDC. It was confirmed that the RS485 communications to the PLCS Failsafe subsystem and AC load subsystem work well after 100Ω terminations were installed to mitigate the suspected high frequency switching noise in the adjacent power lines. As of the date of writing, the 6 turbines have been working without problems for 3000 hours, generating 3kW to 6.6kW of power.

The system has improved the lives of the residents by supplying electricity for lighting and hot water. It has also generated interest from neighboring villages and local government officials because of the improvements over previous rural electrification practices. Additional sites are being planned and effort is being made to further lower the cost of the system.

V. CONCLUSION

In this paper, we described a novel programmable turbine failsafe subsystem that provides significant benefits over previous failsafe implementations. It provides fractional power diversion and automatic back-off for smooth voltage control, and it mitigates the need for manual reset. It was deployed as an integral part of a priority dump load control (PLCS) system to better utilize excess generated power. Due to the availability of the hardware platform, and the stability of software libraries, the Arduino Mega and the Arduino IDE were chosen. Additionally, JSON was chosen as the communications protocol for the ease of programming and human usability.

The system has been installed and work well and brought benefits of electricity and hot water to the villagers, and has attracted the attention of neighboring villages and government officials. Additional pilot systems are planned to drive down the total system cost to ensure broad replication.